\documentclass{article}
\usepackage[utf8]{inputenc}

\usepackage[final, nonatbib]{neurips_2023_ml4ps}
\usepackage[sort&compress,numbers]{natbib}
\title{Echoes in the Noise: Posterior Samples of Faint Galaxy Surface Brightness Profiles with Score-Based Likelihoods and Priors}

\author{%
Alexandre Adam$^{1,2,4}$ \quad Connor Stone$^{1,2,4}$ \quad Connor Bottrell$^{5,6}$ \quad Ronan Legin$^{1,2,4}$   \\
\textbf{Yashar Hezaveh}$^{1,2,3,4,7,8}$ \quad \textbf{Laurence Perreault-Levasseur}$^{1,2,3,4,7,8}$ \\
$^1$Université de Montréal \quad $^2$Ciela Institute \quad $^3$CCA, Flatiron Institute \quad $^4$Mila \\ \ $^5$ICRAR \quad $^6$Kavli IPMU \quad $^7$Trottier Space Institute \quad $^8$Perimeter Institute\\ 
\texttt{\{alexandre.adam,connor.stone,ronan.legin,yashar.hezaveh,}\\\texttt{laurence.perreault.levasseur\}@umontreal.ca}\\
\texttt{connor.bottrel@icrar.org}
\\
}

\usepackage{aasmacros}  % need the style file in the same folder

\usepackage{float}
\usepackage{amsmath, bm, bbm}
\usepackage{amsthm}
\usepackage{amssymb}
\usepackage{mathtools}
\usepackage{physics}
\usepackage{tikz}
\usepackage{xcolor}
\usepackage{enumitem}

\usepackage[frozencache,cachedir=.]{minted} % for code at the end (arxiv submission)

\usepackage{microtype}
\usepackage{graphicx}
\usepackage{subfigure}
\usepackage{booktabs} 
\usepackage{tikz}
\usepackage{listings}
\usepackage{hyperref}
\hypersetup{
    colorlinks,
    linkcolor={red!50!black},
    citecolor={blue!50!black},
    urlcolor={blue!80!blue}
}
\usepackage{url}

\usepackage{listings}
\lstnewenvironment{algolst}[1][]
{
  
  \lstset{#1}
}{}

\newcommand{\jwst}{\textit{JWST}}

\DeclareRobustCommand{\bbone}{\text{\usefont{U}{bbold}{m}{n}1}}

\theoremstyle{plain}

\theoremstyle{definition}

\theoremstyle{remark}

\begin{document}

\maketitle

\begin{abstract}
    Examining the detailed structure of galaxy populations provides valuable insights into their formation and evolution mechanisms.
    Significant barriers to such analysis are the non-trivial noise properties of real astronomical images and the point spread function (PSF) which blurs structure.
    Here we present a framework which combines recent advances in score-based likelihood characterization and diffusion model priors to perform a Bayesian analysis of image deconvolution.
    The method, when applied to minimally processed \emph{Hubble Space Telescope} (\emph{HST}) data, recovers structures which have otherwise only become visible in next-generation \emph{James Webb Space Telescope} (\emph{JWST}) imaging.
\end{abstract}

\section{Introduction}

% Possible alternate intro
A broad diversity of galaxy morphologies are observed in the local universe~\citep{Walmsley2022} which gives insight into the physical mechanisms and processes by which galaxies evolve.
A number of deep imaging campaigns~\citep{Williams1996,Ferguson2000,Giavalisco2004,Beckwith2006,Davis2007,Scoville2007,Windhorst2011,Grogin2011,Koekemoer2011,Brammer2012,Lotz2017} have made it possible to trace morphological evolution over cosmic time~\citep{Conselice2003,vanderWel2014,Tacchella2019,Bluck2019,Masters2020}. 
The most significant barrier to these evolutionary studies is the corruption by complex noise and blurring due to the point spread function (PSF) of a given instrument~\citep{Starck2002}.
%Lower signal-to-noise ratios of more distant sources could mean that inaccuracies in the noise model could result in more significant biases.
Farther objects project onto a smaller area of a detector, which smooths objects based on distance, the very axis on which such analyses would be most interesting.

Since noise and convolution result in a true loss of information, the inverse problem of recovering the surface brightness profile of an object is ill-posed. Within a Bayesian inference framework, solving this challenge raises two important problems.
First, a prior encapsulating the space of all possible galaxy morphologies must be defined; diffusion models are well known to effectively represent such complex priors \citep{Remy2022,Graikos2022,Adam2022}.
Other approaches used in the astronomy literature include: imposing a basis of functions~\citep{Hogbom1974, Michalewicz2023}, regularization~\citep{Lucy1994, Bertero1998, Starck2002}, and deep networks to approximate the deconvolution function~\citep{Dong2020,Akhaury2022} or implicitly represent the prior~\citep{Morningstar2018,Morningstar2019,Adam2022RIM,Wang2023}. Forward modeling with an explicit representation (prior) for the model and a PSF present a statistically principled path forward, though to date this has only been achieved with restricted parametric prescriptions~\citep{Stone2023}.
The second problem is to obtain an accurate characterization of the noise in real astronomical images (the likelihood). %, which subsequently determines the likelihood.
This point is often overlooked in favor of assuming Gaussian uncertainties on pixel fluxes, though it is well known that pixels experience significant non-Gaussian sources of noise such as cosmic rays, non-linear response functions, blooming, and cross-talk (to name a few).
%To tackle this second problem we use recent advances in likelihood approximation using a SLIC model~\citep{Legin2023}, thus for the first time we may approach the galaxy deconvolution problem with a complete Bayesian treatment.
%While the likelihood of pixels is typically assumed to be independent Gaussians, a great body of work has been applied to solving the prior in deconvolution problems.
%This prior may take a number of forms: one may impose a basis of allowed functions~\citep{Hogbom1974, Michalewicz2023}, impose regularization to penalize high-frequency structures~\citep{Bertero1998, Starck2002}, or allow a machine learning algorithm to approximate the deconvolution function~\citep{Dong2020,Akhaury2022}.
%Further methods have been developed, such as splitting an image into two channels~\citep{Lucy1994,Michalewicz2023}, separating the treatment of point sources and diffuse emission, which require alternate prescriptions.
%All of these methods have varying advantages in terms of minimizing artifacts, preserving flux, speed, and the reliability of data products.

 Here we demonstrate a method to build and use an accurate likelihood and prior probability distribution for the problem of inferring the true galaxy images on pixelated grids via forward modelling. %, overcoming the need for ad-hoc heuristics, parametric prescriptions, and regularization. 
%We will also show how combined with a data-driven noise model we can use this prior to generate posterior samples in the high-dimensional space of image pixels. 
We do this by taking advantage of recent advances in score-based models~\citep{Ho2020,song2021sde} in regards to astronomical applications~\citep{Smith2022}, and likelihood modelling~\citep{Legin2023}, to frame deconvolution as a Bayesian inference problem.
We demonstrate the performance of this method by generating posterior samples from observations of galaxies in minimally processed \emph{HST} data to reveal detailed structures in remarkable detail. These structures are then confirmed with \emph{JWST} data of the same sources, showing the effectiveness of the method to recover information that is otherwise indiscernible.

\begin{figure}[H]
    \centering
    \includegraphics[width=\textwidth]{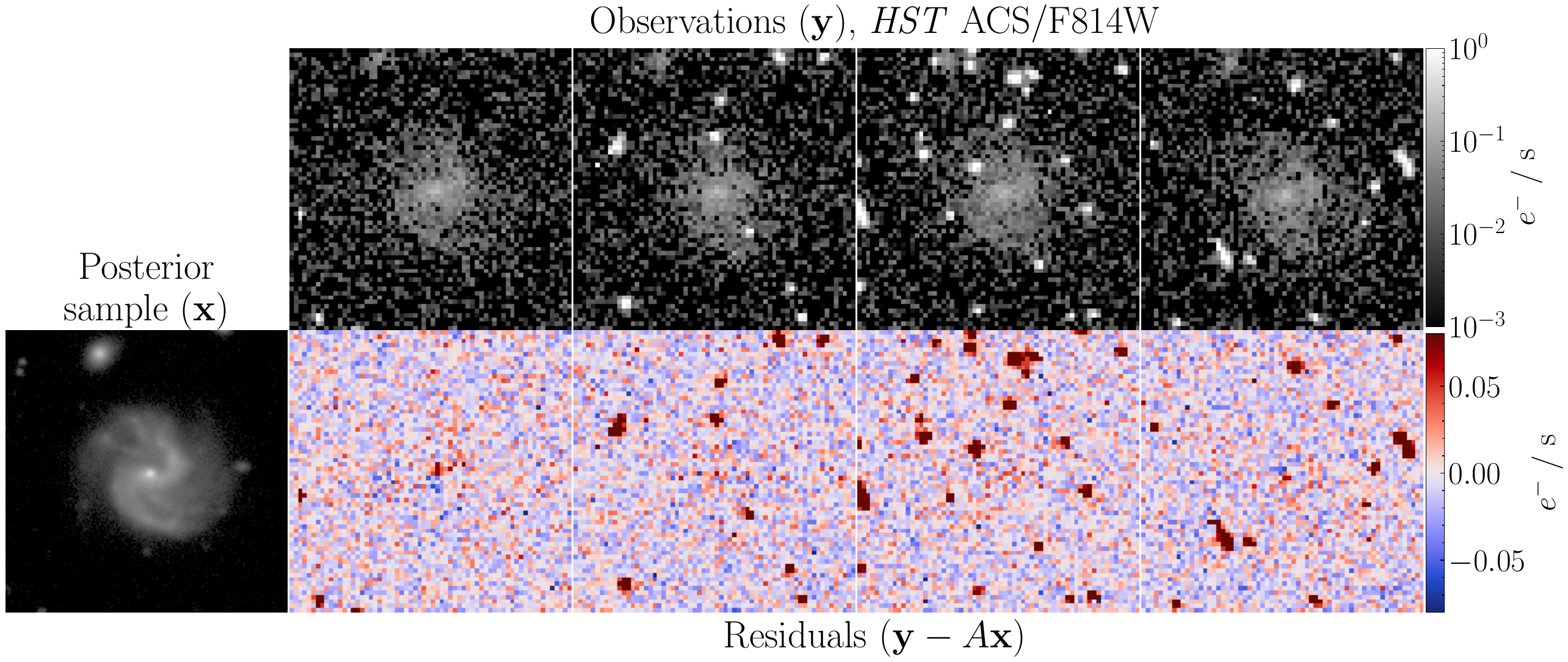}
    \caption{An example of a posterior sample for a COSMOS \citep{Scoville2007} target located at ${\mathrm{R.A.}=9^{\mathrm{h}}57^{\mathrm{m}}56^{\mathrm{s}}.5294}$, ${\mathrm{Dec.}=2^\circ27'37''.963}$. The top row shows the 4 exposures jointly used in the likelihood. The bottom row shows a posterior sample (left) and the corresponding residuals (observation minus posterior sample) for this model.}
    \label{fig:residuals}
\end{figure}

\section{Bayesian inference of galaxy surface brightness blurred with a PSF and in the presence of non-Gaussian noise}

Our goal is to infer the posterior distribution of the surface brightness of galaxies on a pixelated grid. We denote the model galaxy images with $\mathbf{x} \in \mathbb{R}^n$, where $n$ is the number of pixels in the reconstructions, and the noisy telescope data with $\mathbf{y} \in \mathbb{R}^m$, where $m$ is the number of data pixels. The data-generating process can be written as 
\begin{equation}\label{eq:inverse_problem}
    \mathbf{y} = A \mathbf{x} + \boldsymbol{\eta}\, . 
\end{equation}
where, $A \in \mathbb{R}^{m \times n}$ captures the telescope's response process, including the blurring by the aperture --- also known as the PSF --- and wavefront aberrations from the optics. In this equation, $\boldsymbol{\eta} \in \mathbb{R}^m$ is a vector of additive noise, which captures the contribution of \emph{all} additive stochastic phenomena other than the signal of interest to the data (e.g., CCD thermal noise, readout noise, cosmic rays, even astrophysical foreground/backgrounds, etc.). Since the noise is assumed to be additive, the likelihood can be written from the noise distribution directly: $\mathbf{y} - A \mathbf{x} \sim p(\boldsymbol{\eta}) = p(\mathbf{y} \mid \mathbf{x})$.

The posterior distribution over $\mathbf{x}$ (model galaxy images) is proportionately given through Bayes' theorem as the product of the prior $p(\mathbf{x})$ and the likelihood $p(\mathbf{y} \mid \mathbf{x})$
\begin{equation}\label{eq:bayes}
    p(\mathbf{x} \mid \mathbf{y}) \propto p(\mathbf{y} \mid \mathbf{x}) p(\mathbf{x})\, .
\end{equation}
%The likelihood term

An accurate characterization of the prior over the underconstrained variable $\mathbf{x}$ is crucial for this inference, as highlighted in numerous studies \citep{Song2021mri,Adam2022,Chung2022,Feng2023a,Feng2023b}.
However, understanding the telescope noise distribution, $p(\boldsymbol{\eta})$, is of paramount importance to enable unbiased inference on real data.
In the following section, we briefly overview the methodology employed to use data samples to learn both distributions --- the prior and the likelihood --- and combine them in order to sample from a close approximation to the posterior.

% \newpage
\subsection{Prior and likelihood score-based modelling with Stochastic Differential Equations (SDE)}
% \begin{wrapfigure}{r}{0.5\textwidth}
    % \centering
%     \includegraphics[width=0.5\textwidth]{figures/comparison_smacs_target2_F160W_posterior_summaries.pdf}
%     \caption{}
%     \label{fig:comparison}

    % \vspace{-0.5cm}
% \end{wrapfigure}

\begin{figure}[t]
    \centering
    \includegraphics[width=0.9\textwidth]{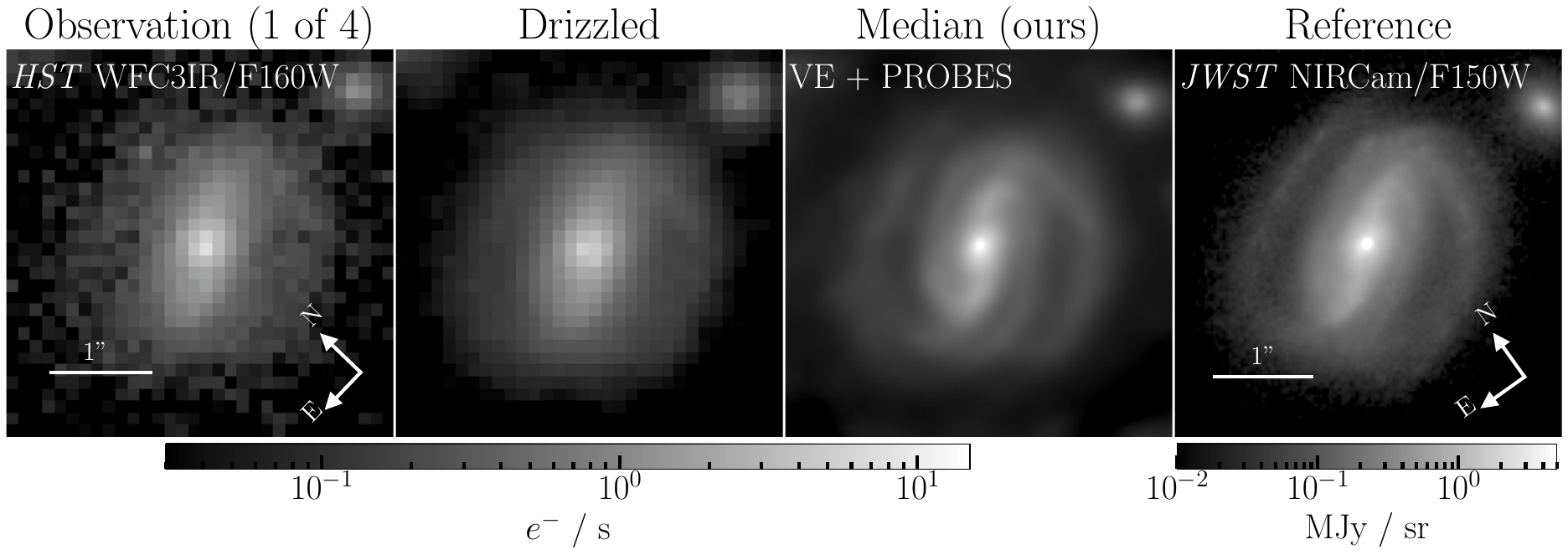}
    \caption{In the leftmost column, we show one of the \emph{HST} exposures in the F160W filter for the target located at ${\mathrm{R.A.}=7^{\mathrm{h}}23^{\mathrm{m}}20^{\mathrm{s}}.7483}$; ${\mathrm{Dec.}=-73^{\circ}26'07''.203}$ in the SMACS 0723 field used to obtain the drizzled image (second column) and the median image composed of $450$ posterior samples using our method (third column). In the rightmost column, we show the corresponding \emph{JWST} image in the closest available filter to \emph{HST}'s F160W filter as a reference. Note that neither the \emph{JWST} image nor the drizzled image were used to perform the inference.}
    \label{fig:comparison}
\end{figure}

We make use of the continuous-time denoising score matching \citep{Hyvarinen2005,Vincent2011,Alain2014} framework developed in \citet{song2021sde} in order to learn the score of the prior, $\grad_\mathbf{x} \log p(\mathbf{x})$, and the score of the noise distribution $\grad_{\boldsymbol{\eta}} \log p(\boldsymbol{\eta})$. 
In summary, the objective function minimizes a weighted sum of Fisher divergences between the output of a neural network with a U-net architecture \citep{Ronneberger2015} and the score of a Gaussian perturbation kernel, $p(\mathbf{x}_t \mid \mathbf{x}_0) = \mathcal{N}(\mathbf{x}_t \mid  \mu(t) \mathbf{x}_0, \sigma^2(t)\bbone_{n \times n})$, both parameterized by the time variable $t \in [0, 1]$ of an SDE. Within this framework, different choices of functions for $\mu(t)$ and $\sigma(t)$ can yield different properties on the generative model \citep[see e.g.][]{song2020improved,Karras2022}. Though any reasonable choice is in principle equivalent, they may differ in numerical stability. In this work, we focus on the well known Variance Exploding (VE) and Variance Preserving (VP) SDEs.  

The strength of the aforementioned framework lies in the simplicity of the perturbation kernel. Assuming that $\mathbf{x}_0$ is a sample from the $t=0$ prior (the distribution of interest: $p(\mathbf{x}_0)$) a sample from the SDE integrated up to time $t$ can be obtained directly from $p(\mathbf{x}_t \mid \mathbf{x}_0)$. This simplicity also allows us to deduce the form of the perturbation kernel for the noise distribution, $p(\boldsymbol{\eta}_t \mid \boldsymbol{\eta}_0)$, aimed at the posterior sampling SDE. Assuming $\mathbf{x}_0$ to be a sample from the posterior, we can use equation \eqref{eq:inverse_problem} and the prior perturbation kernel to obtain
$
    \boldsymbol{\eta}_t = \mu(t) \boldsymbol{\eta}_0 - A \sigma(t) \mathbf{z} ,
$
where $\mathbf{z} \sim \mathcal{N}(0, \bbone_{n \times n})$ and $\boldsymbol{\eta}_0 = \mathbf{y} - A \mathbf{x}_0$. From this equation, we obtain a modified perturbation kernel that accounts for the Brownian motion injected \textit{through} the physical model, $A$, when solving the posterior SDE:
\begin{equation}\label{eq:slic_perturbation_kernel}
    p(\boldsymbol{\eta}_t \mid \boldsymbol{\eta}_0) = \mathcal{N}(\mu(t) \boldsymbol{\eta}_0, \sigma^2(t)AA^T)
\end{equation}
The score of the noise distribution, $\grad_{\boldsymbol{\eta}_t} \log p_t(\boldsymbol{\eta}_t)$, is thus constructed to have more accurate time marginals for the posterior SDE obtained 
by applying Anderson's reverse-time formula \citep{Anderson1982} to the prior SDE and setting the posterior as the $t=0$ boundary condition of the SDE.

Following the work of \citet{Legin2023}, we can recover the likelihood score from the score of the noise distribution by applying the chain rule and using equation \eqref{eq:inverse_problem}
\begin{equation}\label{eq:slic}
    \grad_{\mathbf{x}_t} \log p_t(\mathbf{y} \mid \mathbf{x}_t) \approx -\grad_{\boldsymbol{\eta}_t } \log p_t(\boldsymbol{\eta} _t) A\, .
\end{equation}
However, for time $t \not= 0$, the above equation is an approximation to the true intractable likelihood score \citep{Chung2022,Feng2023a,Feng2023b,Rozet2023}, at $t=0$ it is exact. Our approximation relies on the assumption that the likelihood is much more informative than the prior \citep{Remy2022,Adam2022}. More details are provided in Appendix \ref{sec:A}.

\subsection{Data and Physical Model} \label{sec:data}

The \textbf{PROBES} dataset is a compendium of high-quality local late-type galaxies \citep{Stone2019,Stone2021} that we leverage to learn a prior from empirical observations of galaxies. These galaxies have resolved structures (bars, spiral arms, molecular clouds, etc.), which make them well suited to approximate what a high-resolution galaxy light profile looks like. 
This dataset has also been used in previous studies to train diffusion models \citep{Smith2022,Adam2022}. 
The \textbf{SKIRT TNG} \citep{Bottrell2023} dataset is a large public collection of images spanning $0.3$-$5$ microns made by applying dust radiative transfer post-processing \citep{Camps2020} to galaxies from the TNG cosmological magneto-hydrodynamical simulations\footnote{\url{www.tng-project.org}} \citep{Nelson2019} over redshifts $0.1 \leq z \leq 0.7$. 
While this dataset inherits the assumptions about cosmology and the sub-grid physics that govern the formation of galactic structure in TNG \citep{Weinberger2017,Pillepich2018}, it offers a unique opportunity to build our prior knowledge about the morphology of faint galaxies without interlopers, noise, and measurement effects inherent in the observed PROBES galaxies. % the redshift range $0.1 \leq z \leq 0.7$.
% This synthetic data is simulated for the filters of the Hyper Suprime-Cam Subaru Strategic Program (HSC-SSP) \citep{HSCfilters}. 
In this work, we make use of the available $g$ and $z$ bands in both datasets to train the prior for optical (F814W) and near infrared (F160W) observations respectively. 
% Those bands are shared between the PROBES and the SKIRT TNG dataset, making them ideal for apple-to-apple comparison between the different priors. 

\textbf{The physical model} is built with \texttt{PyTorch} \citep{pytorch} for its automatic differentiation capability. 
We incorporate 2 main components in the forward model, namely \textit{HST}'s optical geometric distortions and the PSF. 
Geometrical distortions are represented through fourth-order polynomial transformations of pixel-to-world coordinates
% \footnote{See the \href{https://hst-docs.stsci.edu/wfc3ihb/appendix-b-geometric-distortion/b-1-overview}{ WFC3 Instrument Handbook} for more details about \textit{HST} geometric distortions.} 
\citep{Bellini2009}. 
In practice, distorted world coordinates are extracted from the FITS files \citep{Pence2010} of each observation using \texttt{WCS} routines from \texttt{Astropy} \cite{astropy:2013,astropy:2018,astropy:2022}. 
A bilinear interpolation is then applied between the model pixel coordinates and the observation world coordinates, similar to what is accomplished with the \texttt{Drizzle} algorithm \citep{Fruchter2002} except without stacking the images. 
For the PSF, we make use of \texttt{PyTorch}'s 2D convolution algorithm and a public library\footnote{\url{www.stsci.edu/hst/instrumentation/wfc3/data-analysis/psf}.}  of 4 times super-sampled PSF models for \textit{HST} \citep{Anderson2006,Bellini2018}. 
Finally, the model resolution is set to match the observation resolution with the average pooling of the pixel values.

\begin{figure}[t]
    \centering
    \includegraphics[width=0.68\textwidth]{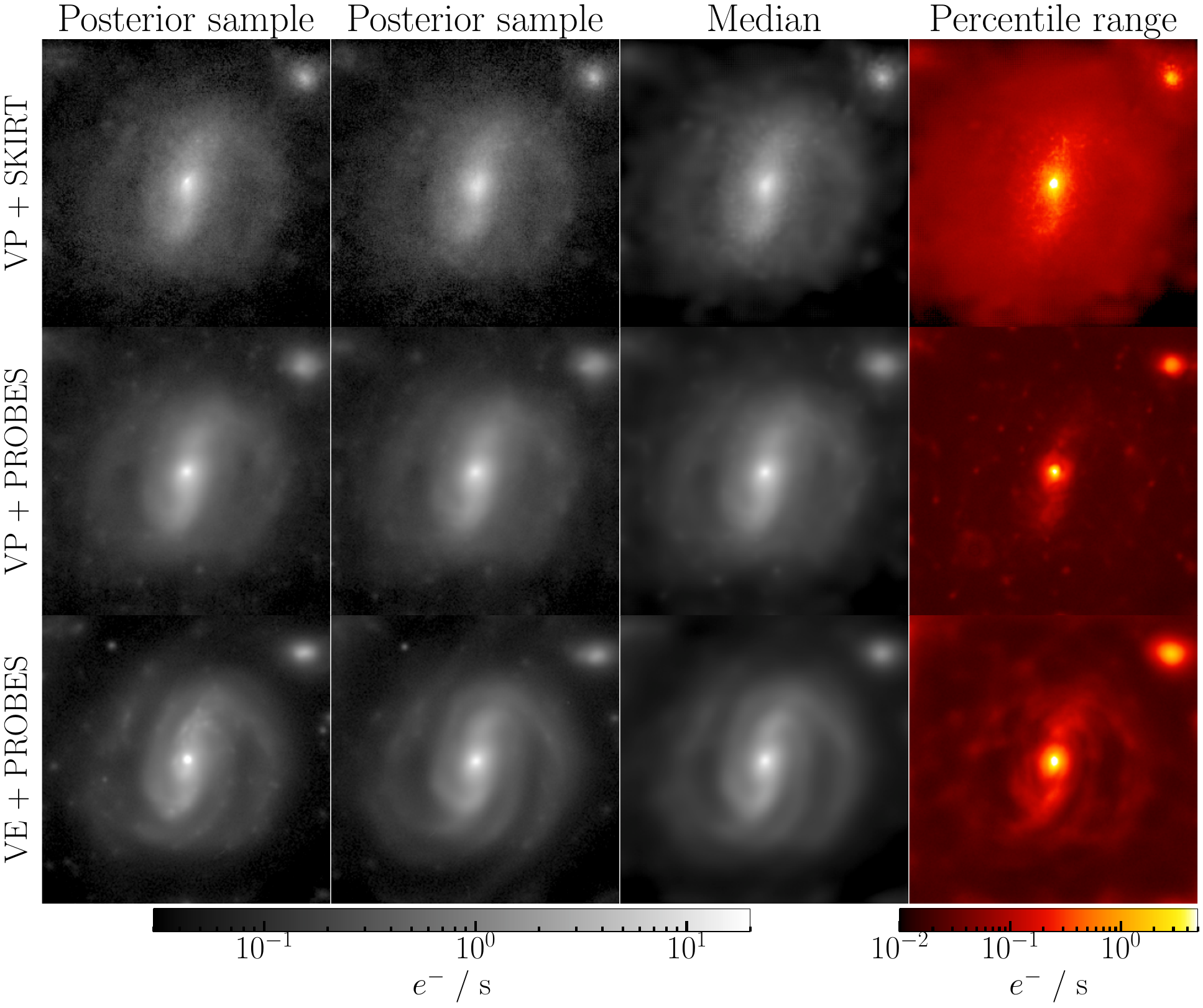}
    \caption{Comparison of the posterior samples, median and percentile range, defined as the $84\%-16\% $ range, for different 
        SDE hyperparameters and prior for the same target as shown in Figure \ref{fig:comparison}. Although the general shape of the image is recovered across the different hyperparameters, our method is able to recover more details using the VE SDE and the PROBES prior.}
    \label{fig:comparison_hyperparams}
\end{figure}

\textbf{The observations} used in this work are taken from two \textit{HST} deep sky fields.
Three observations of faint galaxies were selected from the COSMOS survey \citep{Scoville2007} and three additional targets are in the SMACS 0723 field. The SMACS 0723 targets offer the unique opportunity to compare our reconstructions from \textit{HST} images with corresponding \textit{JWST} images that are both much deeper and higher resolution. 
In this sense, the \textit{JWST} images serve as ground truth for the techniques we deploy in this work. 
For each target, we gather all four dithered exposures captured within a single orbit of the \textit{HST} from the Minkulsky Archive for Space Telescope.
% \footnote{\url{archive.stsci.edu}}. 
More specifically, we use the flat-fielded product from the \textit{HST} pre-processing pipeline. We then multiply each exposure by the pixel area map and subtract the sky background.
Our inference is performed jointly over all the exposures.
% of a target, instead of a single derived \texttt{Drizzle} image.

Finally, to train the score models over samples of the noise distributions, we take random cutouts from the full \textit{HST} exposure for each target based on a flux cut. For the COSMOS targets, we collected 39,254 noise cutouts of shape $64\times 64$ pixels with a flux $F < 0.02\, e^{-1}/s$. For the SMACS 0723 targets, we collected 8,747 noise cutouts of shape $32 \times 32$ pixels with a flux $F < 0.01\, e^{-1}/s$. 
For training, we approximate the physical model, $A$, with a circulant matrix in order to compute the perturbation kernel \eqref{eq:slic_perturbation_kernel} efficiently with a Fast Fourier Transform (FFT) \citep{CooleyFFT1965}. 
More details are provided in Appendix \ref{sec:B}.

\section{Results and Discussion}

As a test that our method can reliably produce samples even in the presence of cosmic rays, we show a reconstructed surface brightness profile in Figure \ref{fig:residuals} at a much higher resolution and depth than \emph{HST} can give us for the F814W filter. Furthermore, the residuals in Figure \ref{fig:residuals} are plausible noise realizations from the telescope noise distribution, showing that all or nearly all of the galaxy surface brightness has been recovered. This is of high value for many morphological studies, which can run automated software on the posterior samples rather than the noisy data. Having access to posterior samples allows one to naturally propagate their uncertainties from the pixel flux measurements to the product of interest.

We validate that our method does not ``hallucinate" morphologies by comparing posterior samples obtained from one of the SMACS 0723 targets with a deeper image of the same target from \emph{JWST} in Figure \ref{fig:comparison}. Note that the JWST image is only shown for reference and was never used during the inference. We also perform reconstructions using different combinations of SDEs and priors. We thus illustrate how different hyperparameter choices can impact the reconstructions, but also showcase how easy it is to swap in a different prior (potentially encoding different underlying assumptions) for a reconstruction.  
% Our framework offers this functionality without modification. 
In principle, the likelihood function can also be easily modified, though should necessarily be tied to the instrument noise statistics. 
As can be seen, important morphological details are recovered in the posterior samples and median which are blurred or buried in noise in the \emph{HST} observation shown in the upper left panel. In particular, the skewed bar feature is recovered, which was mostly blurred out in the \emph{HST} observation. 
The spiral arms are also distinctly recovered. 
The core, however, is not strictly the same as \jwst, using either of our priors. 
Posterior samples of additional targets are presented in Appendix~\ref{sec:C}.

In summary, this work develops a framework for the Bayesian inference of surface brightness of galaxies in the presence of blurring by PSF and non-Gaussian additive noise, allowing maximal extraction of information from noisy data, with a high impact on numerous studies. The results of the method are validated using new higher-quality observations.

% In conclusion, the methods presented in this work have demonstrated their effectiveness at recovering plausible surface brightness profiles of noisy and corrupted \emph{HST} data, pushing the frontiers on what is potentially knowable from the existing pool of \emph{HST} deep fields.

\section{Acknowledgements}
This research was made possible by a generous donation by Eric and Wendy Schmidt with the recommendation of the Schmidt Futures Foundation. 
We are also grateful to Nikolay Malkin for helpful discussions at various points of this project. 
The work is in part supported by computational resources provided by Calcul Quebec and the Digital Research Alliance of Canada. 
C.S. acknowledges support from the Natural Sciences and Engineering Council of Canada (NSERC) and the Canadian Institute for Theoretical Astrophysics (CITA).
Y.H. and L.P. acknowledge support from the Natural Sciences and Engineering Council of Canada grant RGPIN-2020-05073 and 05102, the Fonds de recherche du Québec grant 2022-NC-301305 and 300397, and the Canada Research Chairs Program. 
The work of A.A. and R.L. were partially funded by NSERC CGS D scholarships. 
R.L. acknowledges support from the Centre for Research in Astrophysics of Quebec and the hospitality of the Flatiron Institute. 

Some/all of the data presented in this paper were obtained from the Mikulski Archive for Space Telescopes (MAST). STScI is operated by the Association of Universities for Research in Astronomy, Inc., under NASA contract NAS5-26555. 

Software used: \texttt{astropy} \citep{astropy:2013,astropy:2018}, \texttt{jupyter} \citep{jupyter}, \texttt{matplotlib} \citep{matplotlib} , \texttt{numpy} \citep{numpy}, \texttt{PyTorch} 
 \citep{pytorch}, \texttt{tqdm} \citep{tqdm}, \texttt{pandas} \citep{pandas}

\bibliography{bib}

%%%%%%%%%%%%%%%%%%%%%%%%%%%%%%%%%%%%%%%%%%%%%%%%%%%%%%%%%%%%%%%%%%%%%%%%%%%%%%%
%%%%%%%%%%%%%%%%%%%%%%%%%%%%%%%%%%%%%%%%%%%%%%%%%%%%%%%%%%%%%%%%%%%%%%%%%%%%%%%
% APPENDIX
%%%%%%%%%%%%%%%%%%%%%%%%%%%%%%%%%%%%%%%%%%%%%%%%%%%%%%%%%%%%%%%%%%%%%%%%%%%%%%%
%%%%%%%%%%%%%%%%%%%%%%%%%%%%%%%%%%%%%%%%%%%%%%%%%%%%%%%%%%%%%%%%%%%%%%%%%%%%%%%
\newpage
\appendix
\onecolumn

\section{Background on continuous-time score-matching and the convolved likelihood approximation}
\label{sec:A}
In this appendix, we provide some context to our methods, in particular, equation \eqref{eq:slic_perturbation_kernel} and equation \eqref{eq:slic}. We first give an overview of the continuous-time score-matching framework used to learn the score of the prior, $\grad_{\mathbf{x}_t} \log p_t(\mathbf{x}_t)$, from samples. 
We will then adapt this framework in order to construct an approximation to the likelihood score, $\grad_{\mathbf{x}_t} \log p_t(\mathbf{y} \mid \mathbf{x}_t)$, at every time $t \in [0, 1]$ of the posterior sampling SDE. 
In the following, we will work with stochastic processes defined on a measurable space $(\mathbb{R}^n, \mathcal{B}(\mathbb{R}^n), \{\mathcal{F}_t\}, \mathbf{W})$ where $\mathcal{B}(\mathbb{R}^n)$ is the Borel $\sigma$-algebra associated with $\mathbb{R}^n$, $\{\mathcal{F}_{t}\}$ is a set of filtrations and $\mathbf{W}$ is the Wiener measure. 

\subsection{Learning the prior}
In order to construct a generative model for the prior, $p(\mathbf{x}_0)$, where $\mathbf{x}_0 \in \mathbb{R}^n$ can represent an image of a galaxy, we first construct a perturbation kernel, $p(\mathbf{x}_t \mid \mathbf{x}_0)$, that progressively destroys all the information in available samples from the prior. 
Common choices for this perturbation kernel apply a stochastic process \citep{song2021sde}, $\mathbf{X}_t(\omega): [0, 1] \times \Omega \rightarrow \mathbb{R}^n$, with a dynamic described by an SDE, $d \mathbf{x} = f(\mathbf{x}, t) dt + g(t)d \mathbf{w}$, where $f$ is the drift, $g$ is an homogeneous diffusion coefficient and $\mathbf{w}$ is a Wiener process. 
In this work, we consider the Gaussian perturbation kernel for the Variance-Preserving (VP) or the Variance-Exploding (VE) SDE
\begin{equation}\label{eq:perturbation_kernel}
    p(\mathbf{x}_t \mid \mathbf{x}_0) = \mathcal{N}(\mathbf{x}_t \mid \mu(t) \mathbf{x}_0, \sigma^2(t) \bbone_{n \times n})\, .
\end{equation}
The perturbation kernel is used to train a neural network $\mathbf{s}_\theta(\mathbf{x}, t): \mathbb{R}^n \times [0,1] \rightarrow \mathbb{R}^n$ to approximate the score of the prior, $\grad_{\mathbf{x}_t}\log p(\mathbf{x}_t)$, with a weighted sum of (forward) Fisher divergences as training objective:
\begin{equation}\label{eq:score_matching}
    \mathcal{L}_\theta = 
    \mathbb{E}_{t \sim \mathcal{U}(0,1)}
    \mathbb{E}_{\mathbf{x}_0 \sim p(\mathbf{x}_0)}
    \mathbb{E}_{\mathbf{x}_t \sim p(\mathbf{x}_t \mid \mathbf{x}_0)} 
    \bigg[\lambda(t)\big\lVert 
        \mathbf{s}_\theta(\mathbf{x}_t, t) - \grad_{\mathbf{x}_t}\log p(\mathbf{x}_t \mid \mathbf{x}_0)
    \big\rVert_2^2 \bigg]\, .
\end{equation}
$\lambda(t)$ is a time-dependent weight factor chosen to be equal to the variance of the perturbation kernel, $\lambda(t) \equiv \sigma^2(t)$. See \citet{Song2021likelihood} for a more general derivation of the loss. Since the perturbation kernel \eqref{eq:perturbation_kernel} is Gaussian, we can readily evaluate its score using the reparametrization
\begin{equation}\label{eq:reparametrization}
    \mathbf{x}_t = \mu(t) \mathbf{x}_0 + \sigma(t) \mathbf{z}\, ,
\end{equation}
where $\mathbf{z} \sim \mathcal{N}(0, \bbone_{n \times n})$, such that we can write $\grad_{\mathbf{x}_t}\log p_t(\mathbf{x}_t \mid \mathbf{x}_0) = - \frac{\mathbf{z}}{\sigma(t)}$. The objective \eqref{eq:score_matching} is often written directly in terms of $\pm \mathbf{z}$, which highlights its connection with the denoising autoencoder objective \citep{Vincent2008}. 

\subsection{The convolved likelihood approximation}
The generative model for the prior is constructed by reversing time in the SDE using Anderson's formula \citep{Anderson1982}. In order to sample from the posterior, we must change the $t=0$ boundary condition of the SDE to be the posterior distribution, $p(\mathbf{x}_0 \mid \mathbf{y})$, instead of the prior, which yields the posterior sampling SDE
\begin{equation}\label{eq:posterior_sampling_sde}
    d \mathbf{x} = 
    (f(\mathbf{x}, t) - g^2(t) \grad_{\mathbf{x}} \log p_t(\mathbf{x} \mid \mathbf{y}))dt + 
        g(t) d \bar{\mathbf{w}}\, ,
\end{equation}
where $\bar{\mathbf{w}}$ is a time-reversed Wiener process. By applying Bayes' theorem, we can make use of the score of the prior directly in the posterior sampling SDE without having to retrain or condition the model, $\mathbf{s}_\theta$, on the observation, $\mathbf{y}$:
\begin{equation}\label{eq:Bayes_sde}
    \grad_{\mathbf{x}_t} \log p_t(\mathbf{x}_t \mid \mathbf{y}) = 
    \grad_{\mathbf{x}_t} \log p_t(\mathbf{y} \mid \mathbf{x}_t) + 
    \underbrace{\grad_{\mathbf{x}_t} \log p_t(\mathbf{x}_t)}_{\mathbf{s}_\theta(\mathbf{x}_t, t)}\, .
\end{equation}
Unfortunately, the likelihood in equation \eqref{eq:Bayes_sde} is an intractable quantity to compute since it involves an expectation over $p(\mathbf{x}_0 \mid \mathbf{x}_t)$ of the likelihood at time $t=0$
\begin{equation}
    p_t(\mathbf{y} \mid \mathbf{x}_t) = \int d\mathbf{x}_0\, p(\mathbf{y} \mid \mathbf{x}_0) p(\mathbf{x}_0 \mid \mathbf{x}_t)\, .
\end{equation}
To simplify this expression, we reverse the conditional $p(\mathbf{x}_0 \mid \mathbf{x}_t)$ using Bayes' theorem to get the known perturbation kernel of the SDE
\begin{equation}\label{eq:true_likelihood_2}
    p_t(\mathbf{y} \mid \mathbf{x}_t) = \int d\mathbf{x}_0\, p(\mathbf{y} \mid \mathbf{x}_0) p(\mathbf{x}_t \mid \mathbf{x}_0) \frac{p(\mathbf{x}_0)}{p_t(\mathbf{x}_t)}\, .
\end{equation}
The convolved likelihood approximation consists of taking the ratio $p(\mathbf{x}_0) / p(\mathbf{x}_t)$ to be a constant equal to 1. By ignoring this ratio, the integral becomes a convolution that is tractable, as we will show in the next subsection. 
However, this approximation means that we may introduce some bias in our posterior sampling method. 
We leave it to future work to explore better approximations and their trade-offs, as done e.g., in \citet{Feng2023a,Feng2023b} or \citet{Rozet2023}. 
% It is perhaps worth noting that our approximation converges to the true likelihood at $t=0$, and that it becomes worse as $t \rightarrow 1$. 

\subsection{Learning the likelihood}
We can now construct the perturbation kernel of the noise distribution, ${p(\boldsymbol{\eta}_t \mid \boldsymbol{\eta}_0)}$, where $\boldsymbol{\eta}_0 \in \mathbb{R}^m$ is the additive noise corrupting the observation $\mathbf{y}$ in equation \eqref{eq:inverse_problem}. We multiply by $\mu(t)$ the data generating process, equation \eqref{eq:inverse_problem}, and use the reparametrization of $\mathbf{x}_t$ in equation \eqref{eq:reparametrization} to obtain
\begin{align}
    \nonumber
    \mu(t)\mathbf{y} &= A \mu(t) \mathbf{x}_0 + \mu(t)\boldsymbol{\eta}_0 \\
    \label{eq:eta_t}
\implies \boldsymbol{\eta}_t &= \mu(t) \boldsymbol{\eta}_0 - A \sigma(t) \mathbf{z}\, .
\end{align}
We defined $\boldsymbol{\eta}_t \equiv \mu(t) \mathbf{y} - A \mathbf{x}_t$, the residuals of the model at time-index $t$ of the posterior SDE. This derivation relies on two crucial assumptions, namely that the physical model, $A \in \mathbb{R}^{m \times n}$, is linear, and that the noise in the inverse problem is additive. 
It is also worth noting that the equality $\boldsymbol{\eta}_t = \mu(t) \mathbf{y} - A \mathbf{x}_t$ can be used to infer $\boldsymbol{\eta}_t$ from the model $\mathbf{x}_t$ when solving the posterior SDE. 

We can obtain the distribution of $\boldsymbol{\eta}_t$ from the RHS of equation \eqref{eq:eta_t}. The addition of two random variables results in the convolution of their probability distribution, such that
\begin{equation}\label{eq:convolved_likelihood}
    p_t(\boldsymbol{\eta}_t) = \int d \boldsymbol{\eta}_0\, p(\boldsymbol{\eta}_0) p(\boldsymbol{\eta}_t \mid \boldsymbol{\eta}_0)\, .
\end{equation}
Equation \eqref{eq:convolved_likelihood} is the convolved likelihood approximation of equation \eqref{eq:true_likelihood_2}.
We define the perturbation kernel, $p(\boldsymbol{\eta}_t \mid \boldsymbol{\eta}_0)$, as
\begin{equation*}
    p(\boldsymbol{\eta}_t \mid \boldsymbol{\eta}_0) = p(\boldsymbol{\eta}_t - \mu(t)\boldsymbol{\eta}_0) = p(-A \sigma(t) \mathbf{z})
\end{equation*}
Since $\mathbf{z} \sim \mathcal{N}(0, \bbone_{n \times n})$, we obtain the distribution for the random variable $- A \sigma(t)\mathbf{z}$ by rescaling the covariance of the Gaussian distribution appropriately. Therefore,
\begin{equation}
    p(\boldsymbol{\eta}_t \mid \boldsymbol{\eta}_0) = 
        \mathcal{N}(\boldsymbol{\eta}_t \mid \mu(t) \boldsymbol{\eta}_0, \sigma^2(t)AA^T)\, .
\end{equation}
By using this perturbation kernel in the score-matching objective \eqref{eq:score_matching}, we are able to train a neural network, $s_\phi(\boldsymbol{\eta}, t): \mathbb{R}^m \times [0,1] \rightarrow \mathbb{R}^m$, which can be used to approximate the likelihood in the posterior sampling SDE. We obtain equation \eqref{eq:slic} by applying the chain rule to the approximate likelihood in equation \eqref{eq:convolved_likelihood} following the work of \citet{Legin2023}
\begin{equation*}
    \grad_{\mathbf{x}_t} \log p(\mathbf{y} \mid \mathbf{x}_t) \approx \grad_{\mathbf{x}_t} \log p_t(\boldsymbol{\eta}_t) =  -\underbrace{\grad_{\boldsymbol{\eta}_t} \log p_t(\boldsymbol{\eta}_t)}_{\mathbf{s}_\phi (\boldsymbol{\eta}_t, t)} A\, .
\end{equation*}

In practice, and since we work in high dimensional spaces ($m,n \gtrsim 10^4$), we implement the formula above, also equation \eqref{eq:slic}, known as Score-based Likelihood Characterisation (SLIC) \citep{Legin2023}, with the Vector-Jacobian Product (VJP). To do this, we implement the physical model, $A$, with \texttt{Pytorch} functions instead of using an explicit dense matrix. This strategy trades memory for compute in our sampling algorithm. A sketched implementation is shown in Figure \ref{code:slic}.

% Maybe explain the strategy since we stack observation??

\begin{figure}[h!]
\begin{minted}[frame=lines]{python}
import torch
from torch.func import vjp

def slic_likelihood(y, x, t):
    y_hat, vjp_func = vjp(A, x) # A is a callable function
    eta = mu(t) * y - y_hat # residuals
    score = s_phi(eta, t) # neural network
    return -vjp_func(score)[0]
\end{minted}
\caption{Sketched implementation of the SLIC likelihood with the Vector-Jacobian Product (VJP) operation using \texttt{PyTorch} \cite{pytorch}.}
\label{code:slic}
\end{figure}

\newpage
\section{Circulant physical model approximation for the noise perturbation kernel }
\label{sec:B}

Similarly to the perturbation kernel of the prior, equation \eqref{eq:perturbation_kernel}, we can evaluate the score of the noise perturbation kernel analytically, reparametrized in terms of the random variable $\mathbf{z} \in \mathbb{R}^n$, using equation \eqref{eq:eta_t}
\begin{equation}\label{eq:noise_kernel_score}
    \grad_{\boldsymbol{\eta}_t} \log p(\boldsymbol{\eta}_t \mid \boldsymbol{\eta})
    = - \frac{1}{\sigma(t)} A \mathbf{z} (AA^T)^{-1}\, .
\end{equation}
To be evaluated numerically, the matrix $AA^T \in \mathrm{R}^{m \times m}$ needs to be invertible. In other words, it cannot have null eigenvalues. To ensure this, we add a small number to the diagonal of the matrix $A$. In this work, we add $\lambda=10^{-3}$. 

In order to speed up the training of the noise score models, we approximate the physical model with a circulant matrix. By definition, a circulant matrix is a Toeplitz matrix, but each row is constructed from the same vector with its element cyclically permuted. 
If $A$ would only account for the point spread function, our approximation would be exact. 
Instead, $A \in \mathrm{R}^{m \times n}$ accounts for different transformations, including the different resolution of the model and the observation, which means $A$ is not strictly circulant and is a rectangular matrix with $m < n$. 
Our approximation for training the noise model consists of constructing a circulant matrix from the effective kernel of the physical model
\begin{equation}
    A \approx \mathcal{F}^\dagger (\mathcal{F}\Pi_{k}) \mathcal{F}\, ,
\end{equation}
where we define the effective kernel, $\Pi_{k}$, as the k\textsuperscript{th} column of $A$. 
With this technique, we remove some (but not all) of the null eigenvalues, assuming that most of the energy of the PSF is concentrated within a few pixels. In practice, we extract the column corresponding to the central pixel, $k=m^2/2$, from the physical model by making use of the Jacobian-Vector Product. We sketch an implementation in Figure \ref{code:JVP}. 

With these two approximations, we can evaluate the score of the noise perturbation kernel in linear time using the Fast Fourier Transform \citep{CooleyFFT1965}. We denote the unitary Fourier transform with $\mathcal{F}$ and the Hermitian conjugate with $\dagger$, such that
\begin{equation}
    \grad_{\boldsymbol{\eta}_t} p(\boldsymbol{\eta}_t \mid \boldsymbol{\eta}_0) = - \frac{1}{\sigma(t)} \mathcal{F}^\dagger \mathcal{F}A \mathbf{z} (\mathcal{F}\Pi_k)^{-2}\, .
\end{equation}
This last equation is fast to evaluate since $\mathcal{F}\Pi_k$ is a diagonal matrix in Fourier space. We sketch its implementation in Figure \ref{code:score}.

\begin{figure}[h!]
\begin{minted}[frame=lines]{python}
from torch.func import jvp

def effective_kernel(A, k): 
    x = torch.randn(n) # Random input for A
    v = torch.zeros(n) # vector for the JVP
    v[k] = 1. # Select column k
    return jvp(A, (x,), (v,))[1]
\end{minted}
\caption{Extraction of the column of the linear model implemented as a \texttt{Pytorch} function with the JVP.}
\label{code:JVP}
\end{figure}

\begin{figure}[h!]
\begin{minted}[frame=lines]{python}
power_spectrum = torch.fft.fft2(effective_kernel(A, k)).abs()**2 + 1e-3
def perturbation_kernel_score(z, t): # z is the a noise vector in R^n
    y = torch.fft.fft2(A(z)) # A is the physical model
    score_tilde = y / power_spectrum
    return -torch.fft.ifft2(score_tilde).real / sigma(t)
\end{minted}
\caption{Score of the perturbation kernel computed in linear time.}
\label{code:score}
\end{figure}

\newpage
\section{Additional Figures}
\label{sec:C}
\begin{figure}[H]
    \centering
    \includegraphics[width=\textwidth]{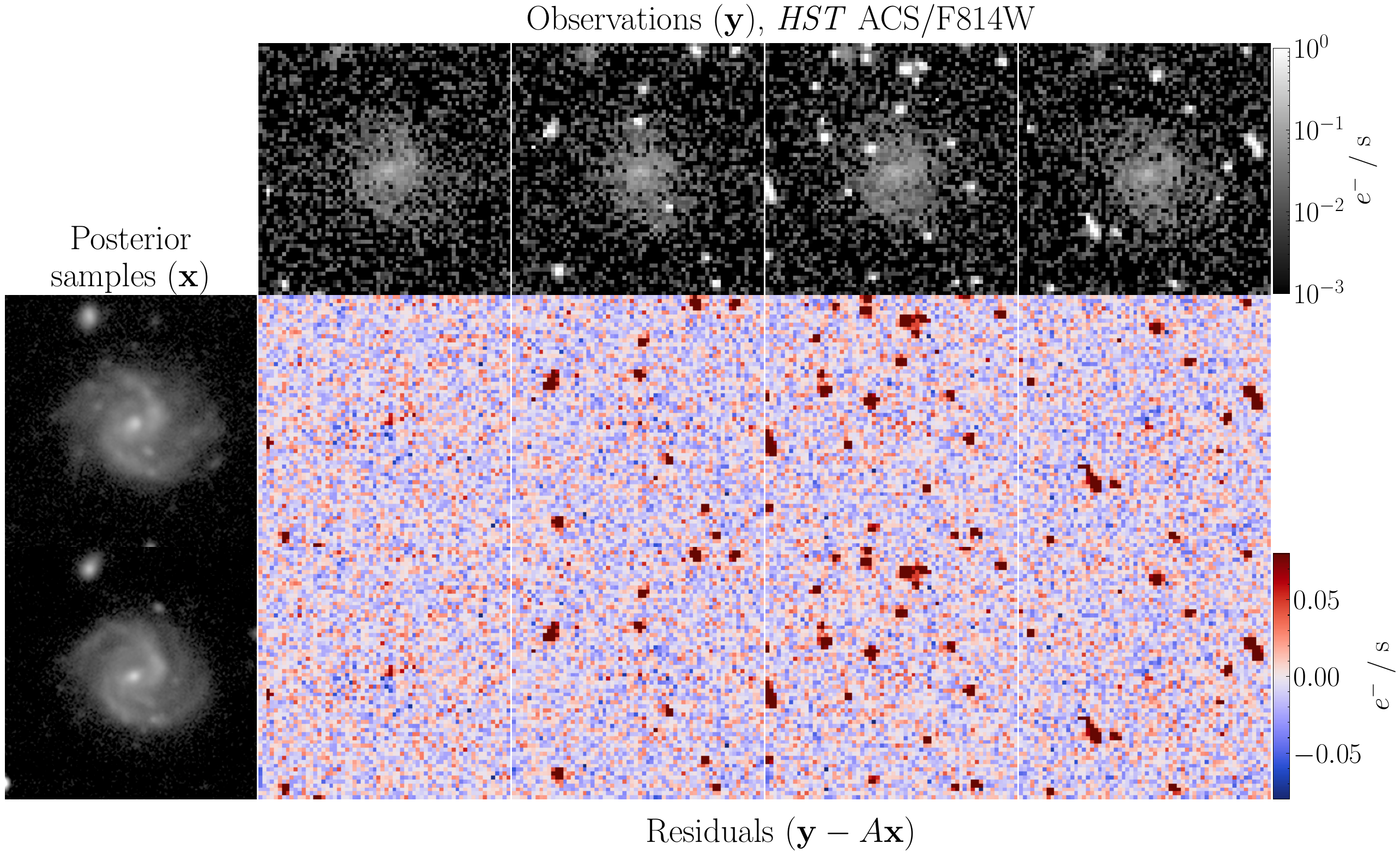}
    \caption{An example of two posterior samples for a COSMOS \citep{Scoville2007} target located at ${\mathrm{R.A.}=9^{\mathrm{h}}57^{\mathrm{m}}56^{\mathrm{s}}.5294}$, ${\mathrm{Dec.}=2^\circ27'37''.963}$. The top row shows the 4 exposures jointly used in the likelihood. The middle and bottom rows show two posterior samples and their corresponding residuals against the observation.}
    \label{fig:residuals1}
\end{figure}

\begin{figure}[H]
    \centering
    \includegraphics[width=\textwidth]{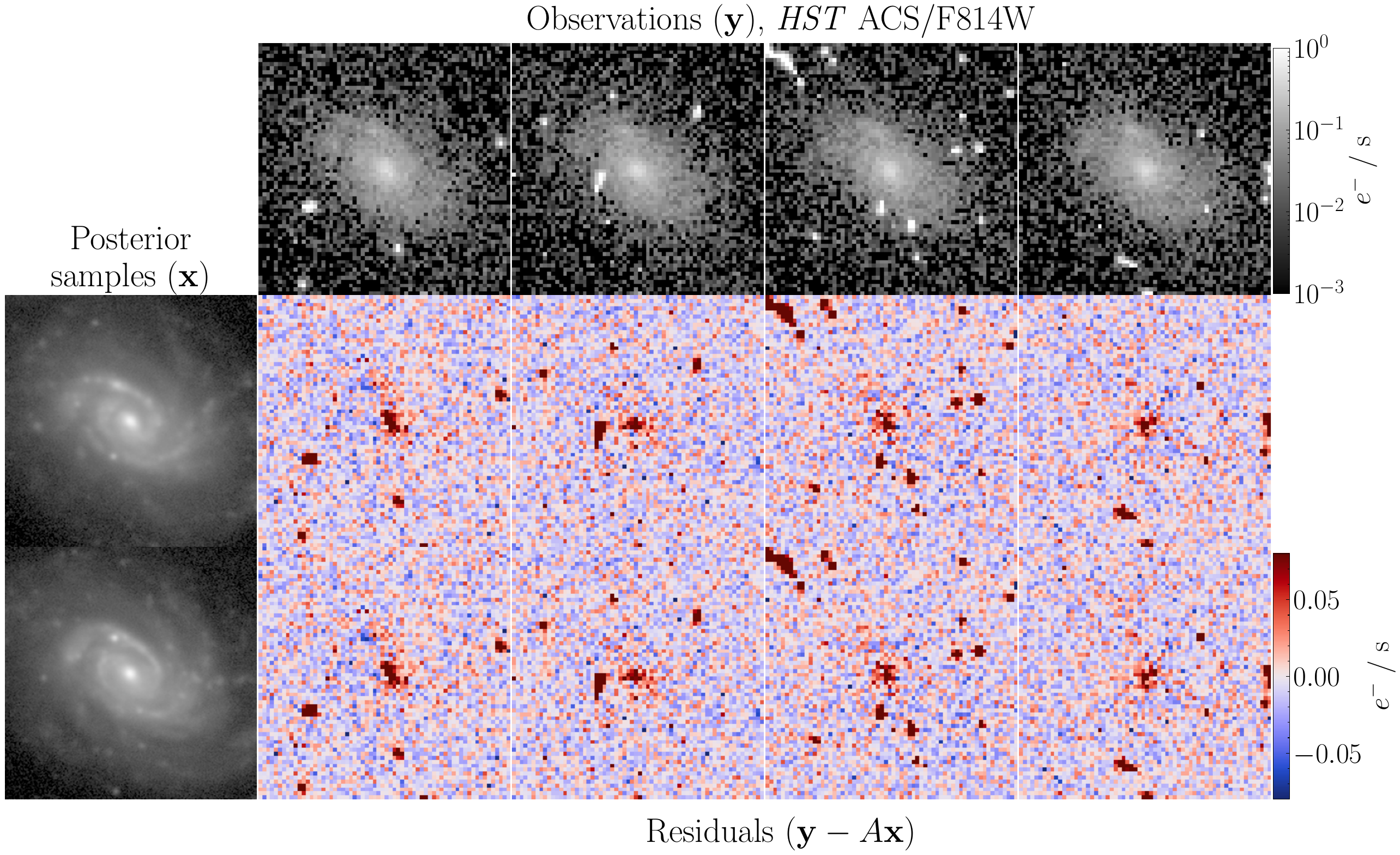}
    \caption{An example of two posterior samples for a COSMOS \citep{Scoville2007} target 
    located at ${\mathrm{R.A.}=9^{\mathrm{h}}57^{\mathrm{m}}46^{\mathrm{s}}.8867}$, ${\mathrm{Dec.}=2^\circ28'22''.735}$. The top row shows the 4 exposures jointly used in the likelihood. The middle and bottom rows show two posterior samples and their corresponding residuals against the observation.
    }
    \label{fig:residuals2}
\end{figure}

\begin{figure}[H]
    \centering
    \includegraphics[width=\textwidth]{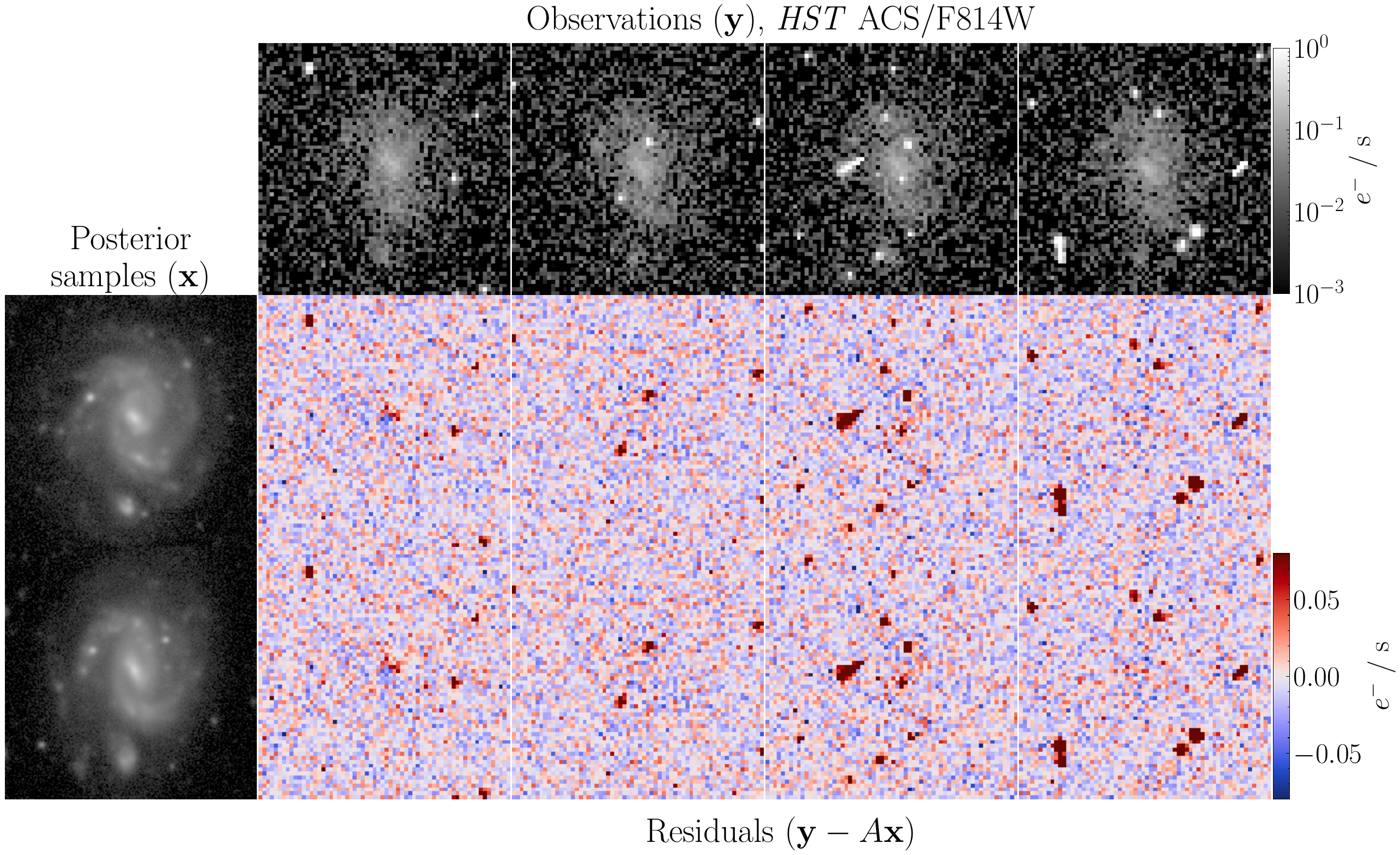}
    \caption{An example of two posterior samples for a COSMOS \citep{Scoville2007} target 
    located at ${\mathrm{R.A.}=9^{\mathrm{h}}57^{\mathrm{m}}49^{\mathrm{s}}.1102}$, ${\mathrm{Dec.}=2^\circ28'19''.430}$. The top row shows the 4 exposures jointly used in the likelihood. The middle and bottom rows show two posterior samples and their corresponding residuals against the observation.
    }
    \label{fig:residuals3 }
\end{figure}

\begin{figure}[H]
    \centering
    \includegraphics[width=\textwidth]{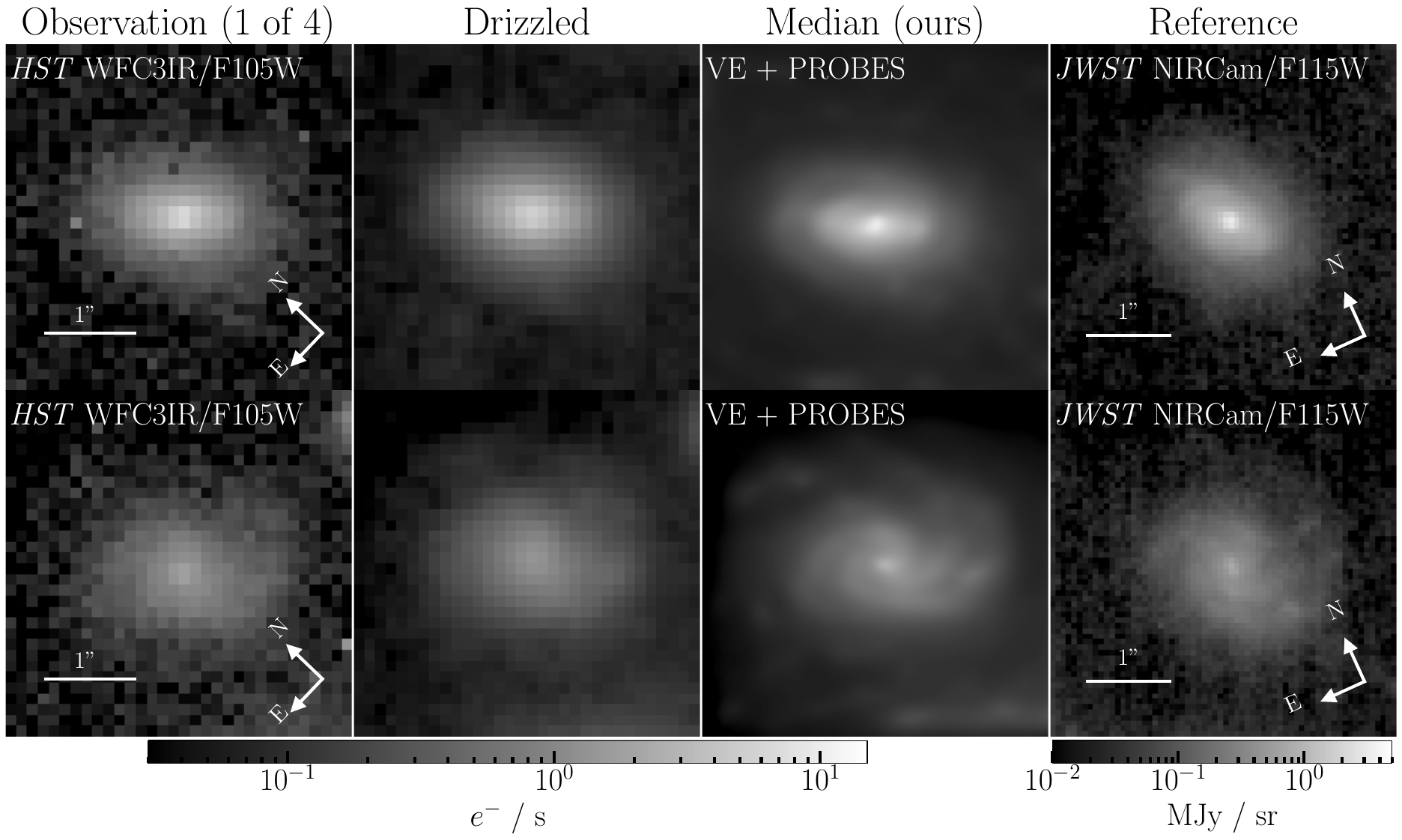}
    \caption{
    In the first row, leftmost column, we show one of the \emph{HST} exposures in the F105W filter for the target located at
    ${\mathrm{R.A.}=7^{\mathrm{h}}23^{\mathrm{m}}24^{\mathrm{s}}.5462}$; ${\mathrm{Dec.}=-73^{\circ}27'20''.172}$ in the SMACS 0723 field 
    used to obtain the drizzled image (second column) and the median image composed of $450$ posterior samples using our method (third column). In the rightmost column, we show the corresponding \emph{JWST} image in the closest available filter to \emph{HST} F105W filter as a reference. Note that neither the \emph{JWST} image nor the drizzled image were used to perform the inference.
    The second row reports the corresponding results for the target located at ${\mathrm{R.A.}=7^{\mathrm{h}}23^{\mathrm{m}}22^{\mathrm{s}}.0805}$; ${\mathrm{Dec.}=-73^{\circ}27'24''.575}$ in the SMACS 0723 field. 
 }
    \label{fig:comparison0}
\end{figure}

\end{document}